\documentclass[journal]{IEEEtran}
\ifCLASSINFOpdf
\else
\fi
\newcommand{\rev}{\textcolor{black}}
\usepackage[font=small,labelsep=period]{caption}
\makeatletter
\usepackage[ruled,linesnumbered]{algorithm2e}
\usepackage{mathrsfs}
\usepackage{amsmath}
\usepackage{graphicx}
\usepackage{amsfonts}
\usepackage{caption} 
\usepackage{color}
\usepackage{multirow}
\usepackage{bm}

\usepackage{url}
\begin{document}
\title{Massive MIMO As an Extreme Learning Machine}

\author{Dawei Gao, Qinghua Guo, \IEEEmembership{Senior Member, IEEE}, and Yonina C. Eldar, \IEEEmembership{Fellow, IEEE}% <-this % stops a space
\thanks{Copyright (c) 2015 IEEE. Personal use of this material is permitted. However, permission to use this material for any other purposes must be obtained from the IEEE by sending a request to pubs-permissions@ieee.org. This work received funding from the Air Force Office of Scientific Research under grant No. FA9550-18-1-0208. Y. C. Eldar's work was also supported by Futurewei and Huawei Technologies.}
%\thanks{Corresponding author: Qinghua Guo.}
\thanks{D. Gao and Q. Guo are with the School of Electrical, Computer and Telecommunications Engineering, University of Wollongong, NSW 2522, Australia (e-mail: dg687@uowmail.edu.au; qguo@uow.edu.au. Corresponding author: Q. Guo.)}
\thanks{Y. C. Eldar is with the Faculty of Math and CS,
Weizmann Institute of Science, Rehovot, Israel (email:
yonina.eldar@weizmann.ac.il).}
}

\maketitle

% in the abstract or keywords

\begin{abstract}
This work shows that a massive multiple-input multiple-output (MIMO) system with low-resolution analog-to-digital converters (ADCs) forms a natural extreme learning machine (ELM). The receive antennas at \rev{the} base station serve as the hidden nodes of the ELM, and the low-resolution ADCs act as the ELM activation function. By adding random biases to the received signals and optimizing the ELM output weights, the system can effectively tackle hardware impairments, such as the nonlinearity of power amplifiers and the low-resolution ADCs. {Moreover, the fast adaptive capability of ELM allows the design of an adaptive receiver to address time-varying 
effects of MIMO channels. Simulations demonstrate the promising performance of the ELM-based receiver compared to conventional receivers in dealing with hardware impairments.} 
%The computational intensive part of the ELM, i.e., the product of its input matrix and its input, is naturally accomplished by the signal superposition in the air. 
%{T}he low-resolution ADCs aid in handling nonlinear impairments, and the most computation-intensive part of the ELM is naturally accomplished by signal transmission and reception. {We \blu{further} extend the ELM receiver to an adaptive one to \blu{address} time varying massive MIMO channels efficiently, thanks to the fast adaptive capability of ELM.}
\end{abstract}

\begin{IEEEkeywords}
Massive MIMO, ELM, signal detection, nonlinear distortion, low-resolution ADC, hardware impairments.
\end{IEEEkeywords}

% \fi
\IEEEpeerreviewmaketitle

\section{Introduction}
\IEEEPARstart{M}{assive} multiple-input multiple-output (MIMO) systems, where the base station (BS) is equipped with a large number of antennas, is a promising technology for 5G and future generation wireless communications \cite{6824752}. However, the requirement \rev{for} a large number of radio frequency (RF) chains leads to high power consumption. {To address this challenge, \rev{various techniques have been proposed to reduce the} number of RF chains \cite{8691587} and low-resolution analog-to-digital converters (ADCs) have been considered \cite{7894211}, \cite{2020arXiv200204290S}, \cite{5351659}. In addition to low-resolution ADCs at the BS, hardware impairments of the user equipment may also need to be addressed. For example, the use of cheap power amplifiers (PAs) can lead to severe nonlinear distortions to transmitted signals \cite{6891254,1264205}.} Previous work has shown that hardware impairments have to be  handled properly to avoid system performance degradation \cite{7894211}, \cite{2020arXiv200204290S}, \cite{5351659}, \cite{1264205}. Existing results consider impairments on either the BS or the user side, however they do not consider impairments on both sides.

\rev{An} extreme learning machine (ELM) is a single-hidden layer feed-forward neural network (NN). In an ELM, the input weights and biases are randomly initialized and fixed, \rev{so that the} only parameters to be learned are its output weights.  \rev{Learning} reduces to solving a least squares (LS) problem, making ELM fast in learning \cite{huang2006extreme}. ELM receivers have been designed in our previous works \cite{2019arXiv190301551G,9040899} to handle LED nonlinearity and/or cross-LED interference. These works show that ELM is very effective in dealing with nonlinear distortions and delivers much better performance than conventional polynomial based techniques \cite{2019arXiv190301551G,9040899}. ELM has also been used for channel estimation and data detection in OFDM systems \cite{8715649,8890830}.

In this work, we address the hardware impairments at both the BS and the user side in a massive MIMO system using the concept of an ELM. {Although deep NNs have attracted much attention for MIMO receiver design (such as \cite{2020arXiv200204290S, 9018199}), we focus on ELM (with a single-hidden layer) due to its fast learning and adaptive \rev{capabilities}. In particular, no back propagation is needed and the size of training samples required is relatively small.} 
We consider the uplink of a massive MIMO system with low-resolution ADCs, where the transmitted signals of users suffer from PA nonlinear distortions. We show that the massive MIMO system forms a \textit{natural} ELM. Specifically, the transmit antennas of users are regarded as the input nodes of the ELM, the massive number of antennas at the BS serve as the hidden nodes of the ELM so that the channel matrix functions as the ELM input weight matrix, and the low-resolution ADCs act as the activation function. We add random biases to the received signals before the ADC and learn the output weights of the ELM from training signals. {As the ELM output weight optimization is simply an LS problem, an adaptive receiver is designed to deal with time-varying channels. Simulation results demonstrate the promising performance of the ELM based system in handling hardware impairments compared to conventional techniques. %The results validate the interesting idea of using low-resolution ADCs to combat PA nonlinear distortions, in the light of ELM.
}

{Notations: Boldface lower-case and upper-case letters denote vectors and matrices, respectively. The superscripts
$(\cdot)^T$ and $(\cdot)^H$ represent the transpose and conjugate transpose operations. We use $|x|$ and $||\bm{x}||$ to denote the amplitude of $x$ and the norm of $\bm{x}$, and $\Re\{\cdot\}$, $\Im\{\cdot\}$ \rev{represent} the real and imaginary parts of a complex number, respectively.}
%The rest of the paper is organized as follows. In Section II, the signal model for massive MIMO with hardware impairments is presented. ELM is briefly introduced in Section III. In Section IV, an ELM receiver is borrowed from \cite{2019arXiv190301551G} for massive MIMO detection. In Section V, the new ELM based receiver is proposed, where the massive MIMO itself is treated as part of the ELM. Simulation results are provided in Section VI, followed by conclusions in Section VII. 

\section{Massive MIMO with Hardware Impairments } 
Consider the uplink transmission of a massive MIMO system with $K$ active users. Each user has a single antenna and the BS is equipped with $N$ antennas, where $N $ can be much larger than $K$. We consider two hardware impairments: PA nonlinear distortion at the transmitter (user) side and low-resolution ADCs at the receiver (BS) side. 

The PA nonlinear distortion can be characterized by the amplitude to amplitude conversion $A(|x|)$ and amplitude to phase conversion $\Phi(|x|)$ \cite{1094911}:
\begin{equation}
    A(|x|)= \frac{\alpha_a|x|}{1+\epsilon_a|x|^2},\quad
  \Phi(|x|)=\frac{\alpha_\phi |x|^2}{1+\epsilon_\phi |x|^2},
\end{equation}
where $x$ is the signal input to the PA, and $\alpha_a$, $\epsilon_a$, $\epsilon_{\phi}$ and $\beta_{\phi}$ are parameters. The distorted signal due to the PA nonlinearity \rev{is} expressed as 
\begin{equation}
    s=f(x)=A(|x|) e^{j(\text{angle}(x)+\Phi(|x|))},
\end{equation}
{where $\text{angle}(x)$ denotes the angle of the complex signal $x$. }
% \begin{figure}
% 	\begin{center}
% 		\includegraphics[width=1.9in]{p1.pdf}\\
% 		\caption{Structure of ELM.}\label{ELM}
% 	\end{center}
% \end{figure}

\begin{figure}
	\begin{center}
		\includegraphics[width=3.2 in]{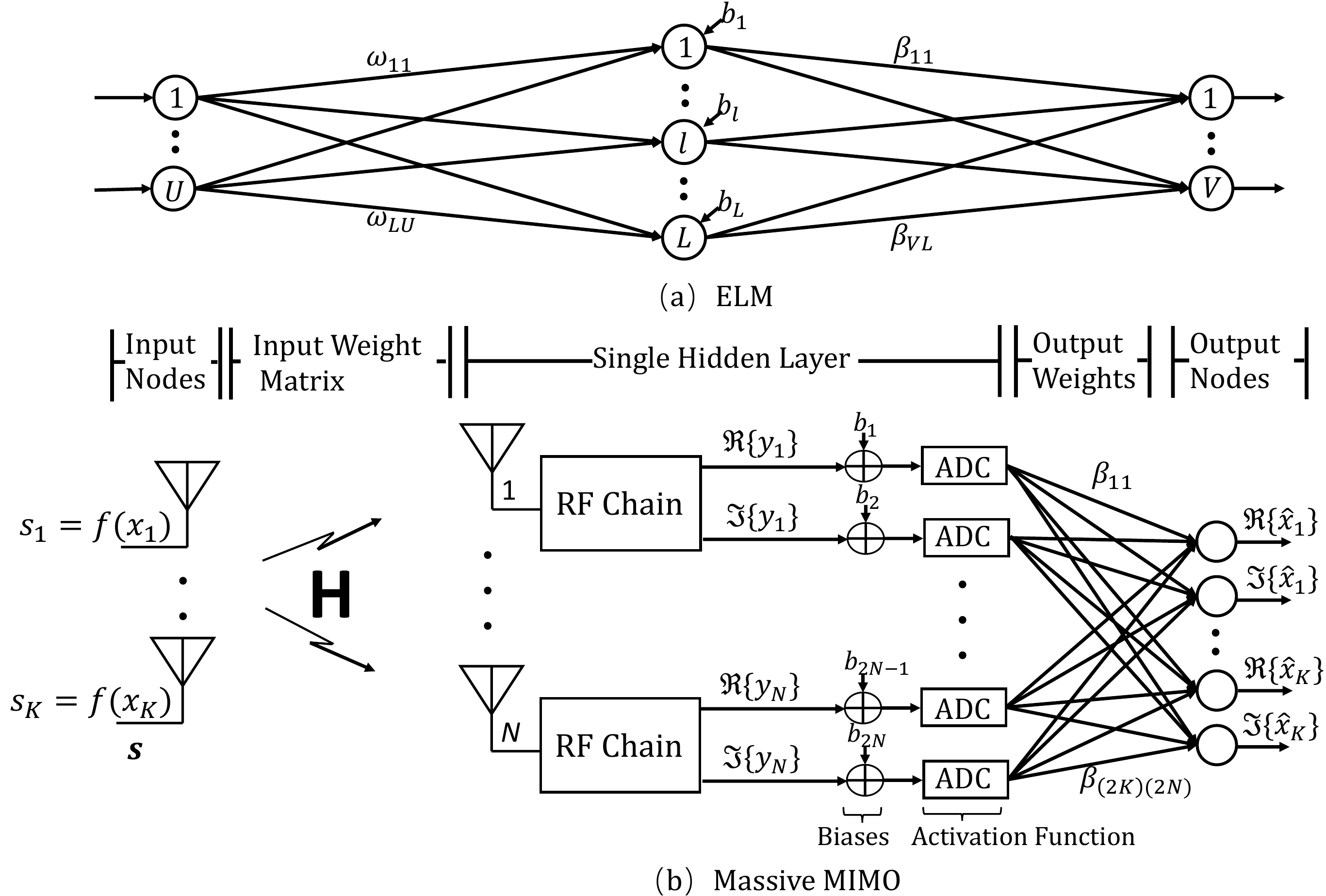}\\
		\caption{{Massive MIMO system forms an ELM, where the received signals are biased before quantization.}}\label{ELM}
	\end{center}
\end{figure}

The received signal at time instant $m$ \rev{is} modeled as
\begin{equation}
    \bm{y}(m)=\bm{H}\bm{s}(m)+\bm{n}(m),
\end{equation}
where $\bm{H}$ is an $N \times K$ MIMO channel matrix, $\bm{s}(m)= f(\bm{x}(m))$ with $\bm{x}(m)= [x_1 (m),x_2 (m),\ldots,x_K (m)]^T$ being the transmitted signals of the $K$ users at time instant $m$, and $\bm{n}(m)$ denotes an additive white Gaussian noise vector. {Conventionally the receiver signal is quantized directly, however in this work, a random bias is added to the received signal, which is then quantized, i.e., 
\begin{align}
    \label{quant}
\bm{r}(m) 
        &= Q(\Re\{\bm{y} (m)\}+\bm{b^{\text{Re}}})+ jQ(\Im\{\bm{y} (m)+ \bm{b^{\text{Im}}}\}),
\end{align}
where $\bm{b^{\text{Re}}}$ and $\bm{b^{\text{Im}}}$ represent the biases for the real part and imaginary part of $\bm{y}(m)$, respectively, and $Q(\cdot)$ denotes an element-wise quantization operation. For a uniform mid-rise quantizer, we have 
\begin{equation}
Q(c)=\Delta \big(\big\lfloor {c}/{\Delta} \rfloor +0.5\big),
\end{equation}
where $c$ denotes the signal to be quantized, $\lfloor\cdot \rfloor$ is the floor function and $\Delta$ is the quantization step size. Due to the limited number of bits used for the ADCs, the input signal $c$ is clipped when its amplitude exceeds a threshold.} The aim of the receiver is to recover $\bm{x}(m)$ based on $\bm{r}(m)$. 

\section{From ELM to ELM Based Massive MIMO System}
%{In this section, we show how a massive MIMO system can be treated as a natural ELM, thereby an interesting receiver can be designed by mimicking ELM.}

\subsection{ELM Structure and Training}

The structure of an ELM with $U$ input nodes, $L$ hidden nodes and $V$ output nodes is shown in Fig. \ref{ELM} (a). The input weights $\{w_{lu}\}$ and biases $\{b_l\}$ are randomly initialized and fixed \cite{huang2006extreme}. The parameters to be learned are the output weights $\{\beta_{vl}\}$.

Suppose we have an input $\bm{p}(m)=[p_{1}(m),p_{2}(m),\ldots,p_{U}(m)]^T$. \rev{The} output of the hidden nodes of the ELM shown in Fig.~\ref{ELM} (a) can be expressed as
\begin{equation}\label{zed}
\bm{z}(m)= g(\bm{W}\bm{p}(m)+\bm{b}),
\end{equation} 
where $\bm{W}$ is the input weight matrix, whose $(l,u)$th element $\omega_{lu}$ associates the $u$th input node and the $l$th hidden node, $g(\cdot)$ is the activation function, and  $\bm{b}=[b_1,b_2,\ldots,b_L]^T$ is the bias vector. The $v$th ELM output \rev{is then} 
\begin{equation}\label{es}
{\psi}_{v}(m) = \bm{z}(m)^T \bm{\beta}_v, 
%\sum_{l=1}^{{L}} {\beta}_{vl} g(\bm{w}_l^T\bm{p}(m)+b_l), \hskip 1pc m = 1,\dotsc,M,
\end{equation}
where
$\bm{\beta}_v=[\beta_{v1},\beta_{v2},\ldots,\beta_{vL}]^T$, and  $\beta_{vl}$ denotes the output weight associating the $l$th hidden node with the $v$th output node.

Suppose that there are $M$ training samples $\{(\bm{p}(m),\bm{t}(m)), m=1,\ldots,M\}$,
%\in\mathbb{R}^U\times \mathbb{R}^V \}_{m=1}^M$, 
where $\bm{t}(m)=[{t}_{1}(m),{t}_{2}(m),\ldots,{t}_{V}(m)]^T$ denotes the expected output. We can \rev{write} (\ref{es}) in matrix form as 
\begin{equation}\label{linear}
\bm{\psi}_v=\bm{Z}\bm{\beta}_v,
\end{equation}
with $\bm{\psi}_v=[{\psi}_{v}(1),{\psi}_{v}(2),\ldots,{\psi}_{v}(M)]^T$, where $\bm{Z}$ is an $M \times L$ matrix given by
\begin{equation}
    \bm{Z}=[\bm{z}(1),\bm{z}(2),\ldots,\bm{z}(M)]^T.
\end{equation}
The output weight vector $\bm{\beta}_v$ \rev{is} trained by minimizing the cost function $\left\lVert \bm{\psi}_v-\bm{t}_v \right\rVert^2$ based on the linear model \eqref{linear}, where $\bm{t}_v=[{t}_{v}(1),{t}_{v}(2),\ldots, {t}_{v}(M)]^T$. The $L_2$ regularized LS solution is given by \cite{4938676}
\begin{equation}\label{beta}
\bm{\beta}_v =(\bm{Z}^T\bm{Z}+\gamma\bm{I})^{-1}\bm{Z}^T\bm{t}_v,\hskip 0.5pc  v=1,...,V
\end{equation}
where $\bm{I}$ is an identity matrix and $\gamma$ is a regularization parameter.

\subsection{Massive MIMO as an ELM}
%\subsection{New ELM Based Massive MIMO}

{Our design is inspired by the fact that the BS is equipped with a large number of antennas $N$ (normally $N>K$), and the quantization is nonlinear. As illustrated in Fig. \ref{ELM}, by biasing the received signals, a massive MIMO system naturally  forms an ELM. } 

%In this section, we treat massive MIMO with low-resolution ADCs as a natural ELM, based on which a new ELM receiver is designed. It is noted that the idea and the receiver are completely different from those in Section IV.

%Figure \ref{s2} illustrates the {new} ELM based massive MIMO system where the transmit antennas, massive MIMO channel and receive antennas serve as part of the ELM. As a common assumption in massive MIMO, we assume that the number of active users $K$ is less than the number of receive antennas $N$ at the base station. 

Comparing Fig. \ref{ELM} (b) to Fig. \ref{ELM} (a), we \rev{consider} the $K$ transmit antennas as the input nodes of the ELM. The distorted signal $\bm{s}(m)=f(\bm{x}(m))$ is transmitted over the air, and detected by the receive antennas at the BS. We treat the receive antennas as the hidden nodes of the ELM, so \rev{that} the channel matrix $\bm{H}$ resembles the input weight matrix of the ELM. The received signals are biased before quantization, as shown in Fig. \ref{ELM} (b). As the signals are complex-valued, we separate their real parts and imaginary parts. {\rev{The} quantized signals \rev{are then} represented in vector form as
\begin{eqnarray}\label{adc}
    	\bm{r}'(m) &=&\! Q\left( \left[ \begin{matrix} \Re\{\bm{H}\bm{s}(m)\} \\ \Im\{\bm{H}\bm{s}(m) \} \end{matrix} \right] \!+\!\bm{b} + \left[ \begin{matrix} \Re\{\bm{n}(m)\} \\ \Im\{\bm{n}(m) \} \end{matrix} \right]\right) \nonumber \\
    	&=& Q\left(\bm{H}'\bm{s}'(m)+\bm{b}+\bm{n}'(m)\right),
\end{eqnarray}
%\! =\!\left[ \begin{matrix} \Re\{\bm{r}(m)\} \\ \Im\{\bm{r}(m) \} \end{matrix} \right] \!
where the length-$2N$ bias vector $\bm{b}$ is randomly generated and fixed,
\begin{equation}
    \bm{H}'=\left[ \begin{matrix} \Re\{\bm{H}\} & -\Im\{\bm{H}\} \\  \Im\{\bm{H}\} & \Re\{\bm{H}\} \end{matrix} \right], ~ 
    \bm{s}'(m) =\left[ \begin{matrix} \Re\{\bm{s}(m)\} \\ \Im\{\bm{s}(m) \} \end{matrix} \right], 
\end{equation}
and $\bm{n}'(m) =[\Re\{\bm{n}(m)^T\}, \Im\{\bm{n}(m)^T\}]^T$.  \rev{Treating} $Q(\cdot)$ as the activation function, the only difference between (\ref{adc}) and (\ref{zed}) is the extra noise term $\bm{n}'(m)$. 

If we ignore the noise term, \rev{then} $\bm{r}'[m]$ resembles the hidden layer output vector of the ELM. By mimicking ELM, the outputs of the ADCs are weighted to recover the signal $\bm{x}(m)$ at the output nodes in Fig. \ref{ELM} (b). \rev{The} transmitters, the massive MIMO channel and the BS receiver form a complete ELM, where the input is the distorted signal $\bm{s}(m)$ and the output is an estimate of $\bm{x}(m)$ (or more precisely estimates of $\Re\{\bm{x}(m)\}, \Im\{\bm{x}(m)\}$).} 

\subsubsection{Receiver training} The ELM output weight vectors $\{\bm{\beta}^{\text{Re}}_k,\bm{\beta}^{\text{Im}}_k, k=1,2,\ldots, K\}$, each pair corresponding to a user, can be learned using training signals $\{ \bm{s}^{\text{Train}}(m), \bm{x}^{\text{Train}}(m), m=1, \ldots, M \}$. {In the resulting ELM, the output vector of the hidden nodes $\bm{r}'(m)$ in \eqref{adc} is formed by signal transmission and low resolution ADC quantization without any additional computations, enabling a low complexity receiver. Collecting the outputs of the ADCs,} we get the hidden node output matrix
\begin{equation}
    \bm{R}^\prime=[\bm{r}^{\prime}(1),\bm{r}^{\prime}(2),\ldots,\bm{r}^{\prime}(M)]^T.
\end{equation}
The values of $\bm{\beta}_k^{\text{Re}}$ and $\bm{\beta}_k^{\text{Im}}$ can then be obtained by solving two regularized LS problems, i.e., 
\begin{equation}\label{betas_spl}
\bm{\beta}_k^{\text{Re}}=({\bm{R}^\prime}^T{\bm{R}^\prime}+\gamma\bm{I})^{-1}{\bm{R}^\prime}^T\bm{t}_k^{\text{Re}},
\end{equation}
\begin{equation}\label{betas_spa}
    \bm{\beta}_k^{\text{Im}}=({\bm{R}^\prime}^T{\bm{R}^\prime}+\gamma\bm{I})^{-1}{\bm{R}^\prime}^T\bm{t}_k^{\text{Im}},
\end{equation}
where
\begin{eqnarray}
\bm{t}_k^\text{Re}= \Re\{[x_k^{\text{Train}}(1),\ldots, x_k^{\text{Train}}(M)]^T \}, \\ \bm{t}_k^\text{Im}= \Im\{[x_k^{\text{Train}}(1),\ldots, x_k^{\text{Train}}(M)]^T \}.
\end{eqnarray}

\subsubsection{Data detection}  Once the output weights for the users are learned, they \rev{are} applied to the received signals to estimate the transmitted data symbols $\{x_k(m)\}$ via
\begin{equation}\label{16}
    \Tilde{x}_k(m)=(\bm{\beta}_k^\text{Re})^T\bm{r}^{\prime}(m)+j(\bm{\beta}_k^\text{Im})^T\bm{r}^{\prime}(m),
\end{equation}
where $\bm{r}'(m)$ is the output of the ADCs at time instant $m$. A decision based on $\Tilde{x}_k(m)$ is then made, i.e.,
\begin{equation}\label{17}
\hat{x}_k(m) = \text{argmin}_c| \Tilde{x}_k(m)-{c}|^2,
\end{equation} 
where $c$ belongs to the symbol alphabet.

\subsection{Comparisons with Other Receivers}
\subsubsection{{Conventional ZF and MMSE receiver}}  

{Given the channel matrix $\bm{H}$, \rev{the} weight for user $k$ \rev{using the zero-forcing (ZF) receiver is} represented as 
\begin{equation}
    	(\bm{w}^{\text{ZF}}_k)^T =\big[(\bm{H}^H\bm{H})^{-1}\bm{H}^H\big]_k,
\end{equation}
where $[\cdot]_k$ denotes the $k$th row of the matrix. If the noise power or the signal to noise power ratio (SNR) is known, we can use the minimum mean squared  error (MMSE) receiver whose weight for user $k$ is 
\begin{equation}
    	(\bm{w}^{\text{MMSE}}_k)^T =\big[\big(\bm{H}^H\bm{H}+\frac{1}{\text {SNR}}\bm{I}\big)^{-1}\bm{H}^H\big]_k.
\end{equation}
Both receivers ignore the nonlinear distortion at the transmitter side and the impact of the low-resolution ADCs at the receiver side, resulting in very poor performance as shown in Section V. It is noted that both receivers require the channel matrix, which needs to \rev{first be} estimated with training signals. }

\subsubsection{ ZF receiver with training} 
The \rev{ZF} receiver \rev{can also directly be} trained using training signals. In this case the weight of the receiver for user $k$  can be expressed as{
\begin{equation}
    	\bm{w}^{\text{ZF}}_{k (\text{Train})} =(\bm{R}^H\bm{R}+\gamma \bm{I})^{-1}\bm{R}^H\bm{x}^{\text{Train}}_k,
\end{equation}
where $\bm{R}=[\bm{r}(1),\bm{r}(2),\ldots,\bm{r}(M)]^T$ is the matrix of the quantized signals (no biasing before quantization), \rev{and} $\gamma$ is a regularization parameter.} {We can also separate the real and imaginary parts, so that the weights \rev{are} obtained \rev{similarly} to \eqref{betas_spl} and \eqref{betas_spa}. \rev{The} difference between the proposed receiver and the trained ZF receiver is that the received signals are biased before quantization in the proposed receiver.} It is interesting that the directly trained detector performs better than the detectors with perfect $\bm{H}$ (see Section V) because the hardware impairments are considered in training.

\subsubsection{ ELM receiver borrowed from {\cite{2019arXiv190301551G}}}

\begin{figure}
	\begin{center}
		\includegraphics[width=3.0in]{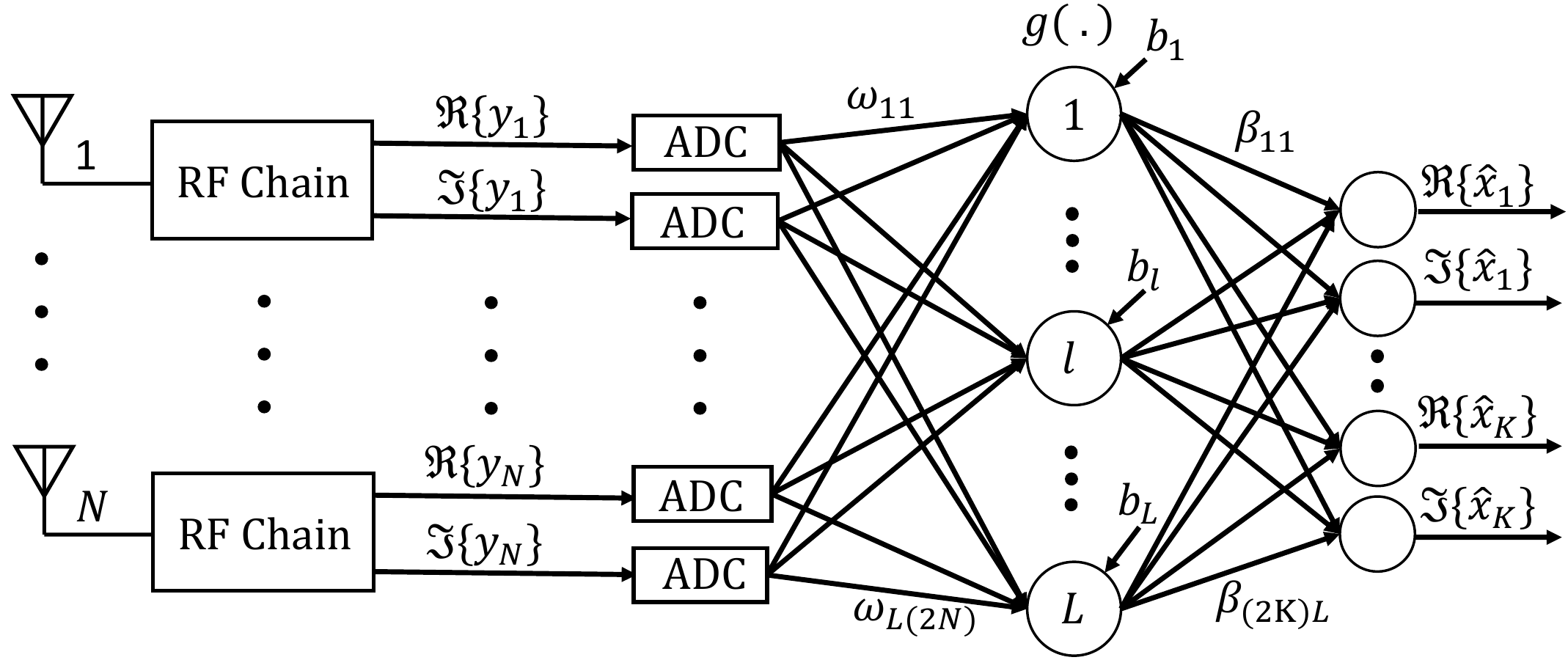}\\
		\caption{ELM receiver borrowed from \cite{2019arXiv190301551G}.}\label{s1}
	\end{center}
\end{figure}

In \cite{2019arXiv190301551G}, we proposed an ELM receiver to handle both the LED nonlinearity and cross-LED interference in MIMO LED communications. The receiver can be readily extended to massive MIMO, as shown in Fig. \ref{s1}. The input to the ELM is the quantized signal (the output of the ADCs), the number of hidden nodes is $L$, and the activation function is the sigmoid function. With the training signals, the output weights of the ELM can also be learned, and then applied for data symbol detection, similar to the proposed receiver.  
%$\{\bm{\beta}^{Re}_k, \bm{\beta}^{Im}_k\}$
%As shown in Fig. \ref{s1}, the  ELM receiver treats the quantized received %signals as the input, and it needs a large number of hidden nodes.
\rev{The} proposed ELM based receiver in Fig. \ref{ELM} (b) is very different, where the multiplication of the input weight matrix with the input vector is naturally accomplished by signal transmission over the air, which leads to much lower complexity of the new ELM based receiver in both training and detection. In data detection, the proposed ELM based receiver only needs to carry out (\ref{16}) and (\ref{17}). However, the ELM receiver borrowed from \cite{2019arXiv190301551G} needs to \rev{implement} matrix-vector multiplications. In addition, as shown in Section V, the proposed ELM based receiver results in considerably better performance.

\section{{ELM Based Adaptive Receiver Design}}

 {ELM is attractive in that it allows fast adaptive learning as its output weights can be easily updated. This endows ELM with the unique capability of dealing with time-varying massive MIMO channels. By leveraging the online sequential-ELM (OSELM) \cite{4012031}, the ELM based receiver \rev{is} readily extended to an adaptive setting.}
 
 %is the online sequential implementation of ELM, where %training data arrives one-by-one or chunk-by-chunk .
 
{The adaptive receiver consists of two learning phases, an initialization phase and a sequential learning phase. Suppose that we have ${M_0}$ training samples for initialization. The output weights for the users can be computed using \eqref{betas_spl} and \eqref {betas_spa}, which are denoted as $\bm{\beta}_k^\text{Re}(0)$ and $\bm{\beta}_k^\text{Im}(0)$ for user $k$. Define
\begin{equation}
\bm{P}(0)=(\bm{R'}^T \bm{R'}+\gamma\bm{I})^{-1}.   
\end{equation}
%{$\{\bm{s}^{\text{Train}}(m),\bm{x}^{\text{Train}}[m], m=1, ...,M_0$, the output weights of the $v$th output node are calculated using (\ref{beta}), i.e.,
%\begin{equation}\label{beta_0}
%\bm{\beta}_v^0 =\bm{P}_0\bm{Z}_0^T\bm{t}_v^0,\hskip 0.5pc  v=1,...,V,
%\end{equation}
%where $\bm{Z}_0=[\bm{z}[1],\bm{z}[2],\ldots,\bm{z}[M_0]]^T$ is the hidden %layer output matrix with the initial training set $\mathcal{M}_0$ as %input, $\bm{P}_0=(\bm{Z}_0^T \bm{Z}_0+\gamma\bm{I})^{-1}$ and  $\bm{t}_v^0=[t_v[1],t_v[2],\ldots,t_v[M_0]]^T$. 
In the subsequent learning phase with new training samples $\{(\bm{s}^\text{Train}(m),\bm{x}^\text{Train}(m))\}$  arriving one by one, the adaptive receiver recursively updates its weights. Taking $\bm{\beta}_k^{\text{Re}}$ as an example, we have 
\begin{equation}\label{k_n}
	\bm{q}(m)=\frac{\bm{P}(m-1)	\bm{r}'(m)}{\lambda+(\bm{r}'(m))^T \bm{P}(m-1)\bm{r}'(m)},
\end{equation}
\begin{equation}\label{e_n}
	{e}_k(m)=\Re\{x_k^{\text{Train}}(m)\}-(\bm{\beta}^{\text{Re}}_k(m-1))^T\bm{r}'(m),	
\end{equation}
\begin{equation}\label{beta_n}
		\bm{\beta}_k^{\text{Re}}(m)=	\bm{\beta}_k^{\text{Re}}(m-1)+\bm{q}(m)e_k(m),
\end{equation}
\begin{equation}	\label{p_n}
	\bm{P}(m)=\frac{1}{\lambda}\big[\bm{P}(m-1)-\bm{q}(m)(\bm{r}'(m))^T\bm{P}(m-1)\big],	
 \end{equation}  
where $\lambda \in [0,1]$ is a forgetting factor to adjust the tracking capability and convergence rate. \rev{The} above sample-by-sample update can also be extended to \rev{a} chunk-by-chunk update \cite{lim2013low}.}

{Although the adaptive receiver may require a relatively large number of training samples (still smaller compared to deep NNs) for initialization, a small number of training samples \rev{are} sufficient for tracking. This makes the adaptive receiver very attractive in handling time-varying massive MIMO channels, as demonstrated in Section V.}

\begin{table}
\caption{{Parameters setting for massive MIMO channel}}\label{tab1}
\begin{center}
\begin{tabular}{ |c|c| } 
\hline
	Parameters of the massive MIMO channel & Value  \\
\hline
Number of antennas $N$ at BS & 256 \\ 
Number of users $K$ & 10 \\ 
Carrier frequency & 2GHz \\ 
Symbol duration $T_s$ & 1$\mu$s \\ 
Mean value of AOA& [$-0.5\pi$,$0.5\pi$] \\
Angular spread (AS)& $10^{\circ}$ \\ 
Number of multiple rays for each user& 5 \\ 
%User mobile velocity & 100 km/h {??????}\\
\hline
\end{tabular}
\end{center}
\end{table}

\section{Simulation Results}

{Assume that the BS is equipped with a uniform linear array (ULA) of $N=256$ antennas, number of users $K$ = 10, and 16-QAM is used. As in \cite{1094911}, the  parameter setting for the PA nonlinearity is  $\alpha_a=1.96$, $\epsilon_a=0.99$, $\alpha_\phi=2.53$ and $\epsilon_\phi=2.82$. Assume 6-bit ADCs. The SNR is defined as $\text{SNR}= P_s/\sigma^2_n$, where $P_s$ is the power of the signal (per transmit antenna), and $\sigma^2_n$ is the power of the noise (per receive antenna). We employ the massive channel model in \cite{8410591},  \cite{1237133} and \cite{7332961}, which considers both spatial and temporal correlations, and assume that the ULA has a half-wavelength spacing. Table \ref{tab1} lists the parameters used for massive MIMO channel generation. Small scale fading is considered and the multiple rays of each user are normalized. The power angular spectrum is modelled using a truncated Laplacian distribution. We consider the channel under a typical mobility scenario, i.e., urban macro \cite{7332961}. For the ZF receiver with training, the weights are computed with real and imaginary parts separated}. For the ELM receiver in \cite{2019arXiv190301551G}, 512 hidden nodes are used, and the input weights and the biases (also the biases of the new ELM based receiver) are drawn independently from a uniform distribution [-0.1, 0.1]. 

\begin{figure}
	\begin{center}
		\includegraphics[width=3.0in]{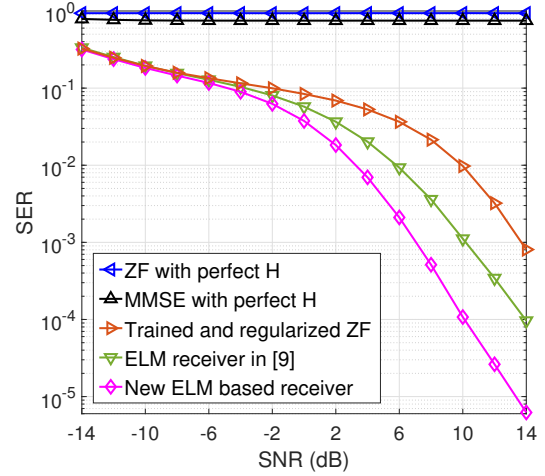}\\
		\caption{{SER performance of various receivers.}}\label{f1}
	\end{center}
\end{figure}
\begin{figure}
	\begin{center}
		\includegraphics[width=3.0in]{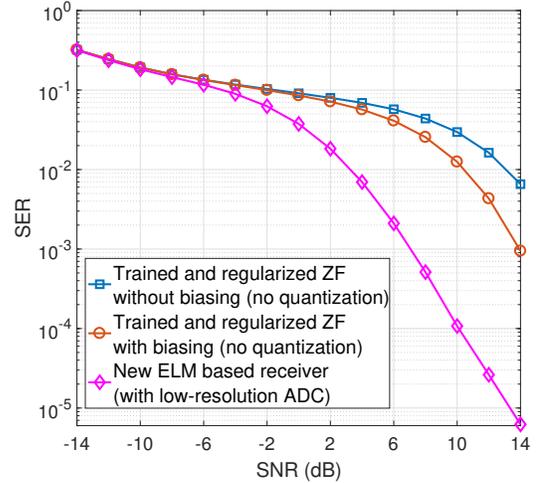}\\
		\caption{{Impacts of biasing and quantization.} }\label{f2}
	\end{center}
\end{figure}

%ADCs with 6-bit quantization are used. The signal-to-noise ratio (SNR) is defined as $SNR= P_s/\sigma^2_n$, where $P_s$ is the power of the signal (per transmit antenna), and $\sigma^2_n$ is the power of the noise (per receive antenna). To train the ELM, ZF and MMSE receivers, training signals with length 3000 are used.

Figure \ref{f1} shows the symbol error rate (SER) of the new ELM based receiver, the ELM receiver borrowed from \cite{2019arXiv190301551G} 
and the ZF and {MMSE receivers} with perfect channel state information and noise power (but note that it is difficult to acquire them when low resolution ADCs are used). Quasi-static channels are assumed without considering the mobility of users, and the training length is 3000. It can be seen from  Fig. \ref{f1} that {the conventional ZF and MMSE receivers deliver poor performance due to the \rev{inability} to mitigate the PA nonlinearity. The trained and regularized ZF receiver performs better than them as the impact of nonlinearity is considered in training. In contrast, the ELM receiver in \cite{2019arXiv190301551G} and the new ELM based receiver can effectively handle the hardware impairments, and the new ELM based receiver delivers the best performance, with significantly lower complexity compared to the ELM receiver in \cite{2019arXiv190301551G}.

The difference between the trained and regularized ZF receiver and the new ELM receiver is that the received signals are biased before quantization in the new receiver. Their huge performance difference indicates the impact of signal biasing.} To further examine the impact of signal biasing and quantization, We assume two trained and regularized ZF receivers without quantization: one performs signal biasing but the other one does not. The results are shown in Fig. \ref{f2}. It is interesting to see that the trained ZF with biasing performs better than the trained ZF without biasing. This indicates that, for a linear receiver, adding biases to the received signals is helpful to deal with the nonlinear distortions. However, the trained ZF receiver with biasing still performs much worse than the new ELM based receiver. This indicates that the low resolution ADCs are helpful in dealing with the nonlinear distortion when they are exploited as the ELM activation function. The ADCs mimic a scaled version of typically used activation functions, e.g., tanh, due to the clipping affect. 

\begin{figure}
	\begin{center}
		\includegraphics[width=3.0in]{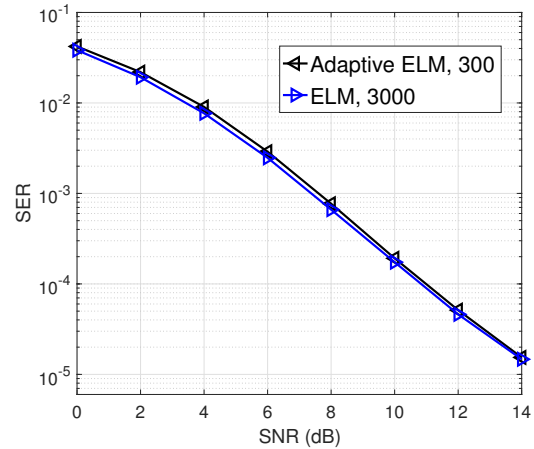}\\
		\caption{{Adaptive receiver with time-varying  channels.}}\label{oselm_300}
	\end{center}
\end{figure}

{Figure \ref{oselm_300} shows the SER performance of the adaptive  receiver with time-varying channels, where a mobile velocity of 100 km/h is assumed. After the initialization with 3000 training symbols, only 300 training symbols in each subsequent frame are used for receiver update. The forgetting factor $\lambda$ is 0.98, and chunk-by-chunk update is used. As a benchmark, we also show the performance of the receiver, where we \rev{assume} that 3000 training symbols are available for training in each frame. We can see that the adaptive receiver with training length 300 can achieve almost the same performance as that with training length 3000, indicating that training length 300 is sufficient for the receiver to track the channel.}

{Note that no explicit channel estimation is needed for the new receiver and the ZF receiver with training, which is in contrast to the conventional ZF and MMSE receivers. Batch training of the new ELM receiver and the ZF receiver requires the same complexity of $\mathcal{O}(N^3+MN^2)$. 
As for detection, the new ELM receiver has the same complexity as the ZF and MMSE receivers, i.e., $\mathcal{O}(N)$ per user.}

%the channel estimation and the computation of the weights of the ZF and MMSE %detectors requires $\mathcal{O}(K^3+KMN+K^2 N+K^2M)$. All the receivers have %the same complexity $\mathcal{O}(N)$ per user for signal detection.}

\section{Conclusion}
We have shown that massive MIMO with low resolution ADCs can be treated as a natural ELM where the massive number of antennas serve as the hidden nodes and the ADCs act as the activation function of the ELM. By adding biases to the received signals and optimizing the ELM output weights, the receiver can effectively handle hardware impairments. {An adaptive receiver \rev{is} also designed and its capability of tracking time-varying channels \rev{is} demonstrated.}

% or


\ifCLASSOPTIONcaptionsoff
  \newpage
\fi



\bibliographystyle{IEEEtran}
\bibliography{bare_jrnl_comsoc.bib}

\begin{thebibliography}{10}
\providecommand{\url}[1]{#1}
\csname url@samestyle\endcsname
\providecommand{\newblock}{\relax}
\providecommand{\bibinfo}[2]{#2}
\providecommand{\BIBentrySTDinterwordspacing}{\spaceskip=0pt\relax}
\providecommand{\BIBentryALTinterwordstretchfactor}{4}
\providecommand{\BIBentryALTinterwordspacing}{\spaceskip=\fontdimen2\font plus
\BIBentryALTinterwordstretchfactor\fontdimen3\font minus
  \fontdimen4\font\relax}
\providecommand{\BIBforeignlanguage}[2]{{%
\expandafter\ifx\csname l@#1\endcsname\relax
\typeout{** WARNING: IEEEtran.bst: No hyphenation pattern has been}%
\typeout{** loaded for the language `#1'. Using the pattern for}%
\typeout{** the default language instead.}%
\else
\language=\csname l@#1\endcsname
\fi
#2}}
\providecommand{\BIBdecl}{\relax}
\BIBdecl

\bibitem{6824752}
J.~G. {Andrews}, S.~{Buzzi}, W.~{Choi}, S.~V. {Hanly}, A.~{Lozano}, A.~C.~K.
  {Soong}, and J.~C. {Zhang}, ``What will {5G} be?'' \emph{{IEEE} J. Sel. Areas
  Commun.}, vol.~32, no.~6, pp. 1065--1082, June 2014.

\bibitem{8691587}
S.~S. {Ioushua} and Y.~C. {Eldar}, ``A family of hybrid analog–digital
  beamforming methods for massive {MIMO} systems,'' \emph{IEEE Trans. Signal
  Process.}, vol.~67, no.~12, pp. 3243--3257, 2019.

\bibitem{7894211}
S.~{Jacobsson}, G.~{Durisi}, M.~{Coldrey}, U.~{Gustavsson}, and C.~{Studer},
  ``Throughput analysis of massive mimo uplink with low-resolution {ADCs},''
  \emph{{IEEE} Trans. Wireless Commun.}, vol.~16, no.~6, pp. 4038--4051, April
  2017.

\bibitem{2020arXiv200204290S}
N.~Shlezinger and Y.~C. Eldar, ``Task-based quantization with application to
  mimo receivers,'' \emph{Commun. Inf. Syst.}, vol.~20, pp. 131--162, 2020.

\bibitem{5351659}
J.~{Singh}, O.~{Dabeer}, and U.~{Madhow}, ``On the limits of communication with
  low-precision analog-to-digital conversion at the receiver,'' \emph{{IEEE}
  Trans. Commun.}, vol.~57, no.~12, pp. 3629--3639, December 2009.

\bibitem{6891254}
E.~{Björnson}, J.~{Hoydis}, M.~{Kountouris}, and M.~{Debbah}, ``Massive {MIMO}
  systems with non-ideal hardware: Energy efficiency, estimation, and capacity
  limits,'' \emph{{IEEE} Trans. Inf. Theory}, vol.~60, no.~11, pp. 7112--7139,
  Nov 2014.

\bibitem{1264205}
L.~{Ding}, G.~T. {Zhou}, D.~R. {Morgan}, {Zhengxiang Ma}, J.~S. {Kenney},
  {Jaehyeong Kim}, and C.~R. {Giardina}, ``A robust digital baseband
  predistorter constructed using memory polynomials,'' \emph{{IEEE} Trans.
  Commun.}, vol.~52, no.~1, pp. 159--165, March 2004.

\bibitem{huang2006extreme}
G.-B. Huang, Q.-Y. Zhu, and C.-K. Siew, ``Extreme learning machine: theory and
  applications,'' \emph{Neurocomputing}, vol.~70, no.~1, pp. 489--501, Dec.
  2006.

\bibitem{2019arXiv190301551G}
D.~Gao and Q.~Guo, ``Extreme learning machine-based receiver for {MIMO} {LED}
  communications,'' \emph{Digit. Signal Process.}, vol.~95, p. 102594, Dec.
  2019.

\bibitem{9040899}
D.~Gao, Q.~Guo, J.~Tong, N.~Wu, J.~Xi, and Y.~Yu,
  ``Extreme-learning-machine-based noniterative and iterative nonlinearity
  mitigation for {LED} communication systems,'' \emph{IEEE Syst. J.}, pp.
  1--10, March 2020.

\bibitem{8715649}
J.~{Liu}, K.~{Mei}, X.~{Zhang}, D.~{Ma}, and J.~{Wei}, ``Online extreme
  learning machine-based channel estimation and equalization for {OFDM}
  systems,'' \emph{{IEEE} Commun. Lett.}, vol.~23, no.~7, pp. 1276--1279, 2019.

\bibitem{8890830}
L.~{Yang}, Q.~{Zhao}, and Y.~{Jing}, ``Channel equalization and detection with
  {ELM}-based regressors for {OFDM} systems,'' \emph{{IEEE} Commun. Lett.},
  vol.~24, no.~1, pp. 86--89, Nov. 2020.

\bibitem{9018199}
H.~{He}, C.~{Wen}, S.~{Jin}, and G.~Y. {Li}, ``Model-driven deep learning for
  {MIMO} detection,'' \emph{IEEE Trans. Signal Process.}, vol.~68, pp.
  1702--1715, 2020.

\bibitem{1094911}
A.~A.~M. {Saleh}, ``Frequency-independent and frequency-dependent nonlinear
  models of twt amplifiers,'' \emph{{IEEE} Trans. Commun.}, vol.~29, no.~11,
  pp. 1715--1720, Nov. 1981.

\bibitem{4938676}
W.~Deng, Q.~Zheng, and L.~Chen, ``Regularized extreme learning machine,'' in
  \emph{Proc. IEEE Symp. CIDM}, March 2009, pp. 389--395.

\bibitem{4012031}
N.~{Liang}, G.~{Huang}, P.~{Saratchandran}, and N.~{Sundararajan}, ``A fast and
  accurate online sequential learning algorithm for feedforward networks,''
  \emph{IEEE Trans. Neural Netw.}, vol.~17, no.~6, pp. 1411--1423, 2006.

\bibitem{lim2013low}
J.-S. Lim, S.~Lee, and H.-S. Pang, ``Low complexity adaptive forgetting factor
  for online sequential extreme learning machine ({OS-ELM}) for application to
  nonstationary system estimations,'' \emph{Neural Comput. Appl.}, vol.~22, no.
  3-4, pp. 569--576, 2013.

\bibitem{8410591}
J.~{Ma}, S.~{Zhang}, H.~{Li}, F.~{Gao}, and S.~{Jin}, ``Sparse bayesian
  learning for the time-varying massive {MIMO} channels: Acquisition and
  tracking,'' \emph{IEEE Trans. Wirel. Commun.}, vol.~67, no.~3, pp.
  1925--1938, 2019.

\bibitem{1237133}
{Ke Liu}, V.~{Raghavan}, and A.~M. {Sayeed}, ``Capacity scaling and spectral
  efficiency in wide-band correlated {MIMO} channels,'' \emph{IEEE Trans. Inf.
  Theory}, vol.~49, no.~10, pp. 2504--2526, 2003.

\bibitem{7332961}
L.~{You}, X.~{Gao}, A.~L. {Swindlehurst}, and W.~{Zhong}, ``Channel acquisition
  for massive {MIMO-OFDM} with adjustable phase shift pilots,'' \emph{IEEE
  Trans. Signal Process.}, vol.~64, no.~6, pp. 1461--1476, 2016.

\end{thebibliography}
\end{document}